\documentclass[aps,prl,twocolumn,superscriptaddress]{revtex4}
\usepackage{graphicx }

\begin{document}

\title{Measurement of the infrared transmission through a single
doped GaAs quantum well in an external magnetic field: Evidence
for polaron effects }

\author{C. Faugeras}
\affiliation{Grenoble High Magnetic Field laboratory,  CNRS, B.P.
166, 38042 Grenoble Cedex 9, France}
\author{M. Orlita}
\affiliation{Grenoble High Magnetic Field laboratory,  CNRS, B.P.
166, 38042 Grenoble Cedex 9, France}

\affiliation  {Institute of Physics, Charles University, Ke
Karlovu 5, CZ-121 16 Praha 2, Czech Republic}
\author{S. Deutchlander}
\affiliation{Grenoble High Magnetic Field laboratory,  CNRS, B.P.
166, 38042 Grenoble Cedex 9, France}

\author{G. Martinez}
\affiliation{Grenoble High Magnetic Field laboratory,  CNRS, B.P.
166, 38042 Grenoble Cedex 9, France}
\author{P.Y. Yu}
\affiliation{Department of Physics, University of California at
Berkeley, Berkeley, CA 94720}

\author{A. Riedel}
\affiliation{Paul Drude Institute, Hausvogteiplatz 5-7, D-10117 Berlin, Germany}
\author{ R. Hey }
\affiliation{Paul Drude Institute, Hausvogteiplatz 5-7, D-10117 Berlin, Germany}
\author{ K. J. Friedland}
\affiliation{Paul Drude Institute, Hausvogteiplatz 5-7, D-10117 Berlin, Germany}

\date{\today}

\begin{abstract}
Precise absolute far-infra-red magneto-transmission experiments
have been performed in magnetic fields up to 33~T on a series of
single GaAs quantum wells doped at different levels. The
transmission spectra  have been simulated with a multilayer
dielectric model. The imaginary part of the optical response
function which reveals new singular features  related to the
electron-phonon interactions has been extracted. In addition to
the expected polaronic effects due to the longitudinal optical
(LO) phonon of GaAs, a new kind of carrier concentration dependent
interaction with interface phonons is observed. A simple physical
model is used to try to quantify these interactions and explore
their origin.

\end{abstract}


\maketitle

Polaronic effects, due the Fr\"{o}hlich interaction between
electrons and the longitudinal optical (LO) phonon of a polar
semiconductor, have been the object of many reports
\cite{Devreese,Gaal,Chen}. In quasi-two-dimensional (Q2D) GaAs
based structures, there is evidence of free polaronic effects (PE)
but few experiments \cite{Wang,Faugeras} have focused on cyclotron
resonance (CR) in the region  above the Reststrahlen Band energy
(RBE) of GaAs (this requires for GaAs a magnetic field strength
beyond 23T). Theoretically  PE were first studied by Lee and Pines
\cite {Lee} and later by Feynman \cite{Feynman}. It was realized
that this kind of interaction could not be properly handled by
perturbation theory and requires a global treatment. Such an
approach was proposed by Feynman \textit{et al.} \cite {FHIP}, and
referred to as the FHIP model. It has been invoked  to explain the
polaronic mass and later extended by Peeters and Devreese \cite
{PD} to extract the conductivity of the Q2D electron gas in the
presence of PE. In the latter case, its major effects are expected
to be observed in the imaginary part of the response function at
energies larger than the RBE. This prediction has motivated the
present study.

Far infra-red magneto-optical experiments have been performed at
magnetic field strengths up to 33~T and at a fixed temperature of
1.8 K, on a series of single modulation-doped quantum wells (QW)
of width $L=13$~nm with different doping levels $N_{s}$ ranging
from 2 to $7.7\times 10^{11}$~cm$^{-2}$ and mobility exceeding
$10^{6}$~cm$^{2}/(V.s)$. The structure of these samples is similar
to those reported earlier \cite {Faugeras}: a single GaAs QW is
sandwiched between two GaAs-AlAs superlattices  $\delta$-doped
symmetrically with  Si-n type dopant on both sides of the QW. In
the present case however the epilayer is not lifted-off from the
GaAs substrate. As a result the samples are optically opaque in
the RBE range. For each fixed value of the magnetic field $B$, an
\textit{absolute} magneto infra-red transmission spectrum
(TA($B,\omega$)) is measured by using a rotating sample holder
containing a hole to obtain a reference spectrum under the same
conditions as for the sample. The Faraday configuration is used
with the \textbf{k} vector of the incoming light parallel to
\textbf{B} and also to the growth axis (100) of the sample. These
spectra TA($B,\omega$) are in turn divided by TA($0,\omega$) to
obtain the
 \textit{relative} transmission spectra TR($B,\omega$) which will be displayed in
the present paper. The analysis of these spectra is based on a
multi-layer dielectric  model \cite{Bychkov}. This is essential
because, in the frequency range of interest, the spectra can be
distorted by dielectric interference effects even in the absence
of electron-phonon interactions.

\begin{figure*}
    \begin{minipage}{0.68\linewidth}
      \scalebox{0.65}{\includegraphics{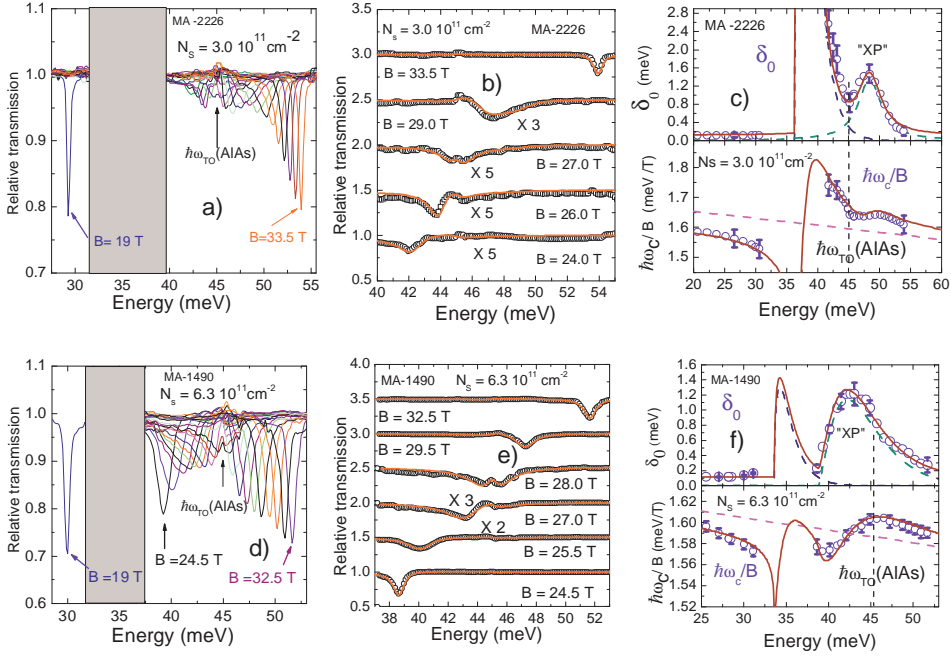}}
    \end{minipage}\hfill
    \begin{minipage}{0.28\linewidth}
      \caption{\label{Fig1} (Color on line) (a) and (d): Relative
transmission spectra of samples MA-2226 and MA-1490 respectively,
for different magnetic fields. (b) and (e): simulation of some
experimental spectra (open dots) with the multidielectric model (
continuous curve): each curve is shifted by 0.5 with a scale
enlarged as noted. (c) and (f) Fitted parameters
$\hbar\omega_{c}/B$ and $\delta_{0}$ as extracted from Eq.~(1) (open
dots) for both samples; the dashed lines in the upper panels
($\delta_{0}$) are the resulting decomposition of the polaronic
and XP interactions adopted to fit $\mathrm{Im}(\Sigma)$; the total
contributions of the interaction (continuous curves in (c) and
(f)) are used to simulate the spectra in (b) and (e) respectively;
the oblique dashed lines in the lower panels of (c) and (f) mimic
the non-parabolicity contribution with the corresponding energy
scale (see text)}
    \end{minipage}
\end{figure*}

The TR spectra of two characteristic samples, sample MA-2226
($N_{s}= 3\times 10^{11}$~cm$^{-2}$) and sample MA-1490 ($N_{s}=
6.3\times 10^{11}$~cm$^{-2}$), are displayed in Figs.~\ref{Fig1}a
and \ref{Fig1}d respectively. For sample MA-2226, the CR
absorption almost vanishes just above the RBE for $B\simeq 24$~T.
But as $B$ is increased beyond 24 T it starts to increase and also
its lineshape changes with field. A "singularity" appears to occur
when the CR energy approaches 45~meV (the transverse optical
phonon (TO) energy of AlAs as measured in the TA(0) spectrum)  at
$B\simeq 27$~T. Its linewidth peaks  when its energy is around
48~meV ($B\simeq 29$~T) before recovering its original line shape
below the RBE at even higher fields. In contrast, for sample
MA-1490, the CR transition is clearly seen when it first emerges
above the RBE, its lineshape has broadened already when its energy
is around 38~meV ($B\simeq 24.5$~T), it shows some singular
behavior at 45~meV ($B\simeq 28$~T) while its linewidth appears to
reach a maximum at the same time. Finally it recovers the original
low fields lineshape for $B> 32$~T.

To simulate these complex spectra, we start with the dielectric
function $\overline{\varepsilon}$ of the doped QW in addition to
the appropriate dielectric functions of  undoped barrier layers
\cite{Bychkov}. For the doped QW, the diagonal part
$\varepsilon_{xx}$ is written as:

\begin{equation}
\varepsilon_{xx}= \varepsilon_{L}-\frac{\omega_{p}^{2}}{\omega
[\omega-(\omega_{NP}-\mathrm{Re}(\Sigma))+\imath(\eta+\mathrm{Im}(\Sigma)]}
\end{equation}
in which the $x$-axis is assumed to lie in the plane of the QW,
$\varepsilon_{L}$ is the contribution of the GaAs lattice,
 $\omega_{NP}$ is the contribution to the CR frequency of the
non-parabolicity (NP) effects \cite{Hermann}, $\eta$ is the field
independent contribution of the background defects to the line
width  while $\Sigma(\omega)$ represents the self energy
correction due to interactions  between the electron gas and
elementary excitations to be discussed further. The effective mass
$m^{*}$ entering the plasma frequency $\omega_{p}$ is calculated
with the same NP model. We note that the lineshape and intensity
of the CR transitions at low  and high fields are practically the
same. Hence we can assume that there is no loss of carrier density
during the magnetic field runs.  In the  fitting process, we
\textit{first} neglect the frequency dependence of $\Sigma$ and
are left with two independent parameters $\omega_{c}=
\omega_{NP}-\mathrm{Re}(\Sigma)$ and
$\delta_{0}=\eta+\mathrm{Im}(\Sigma)$ for each magnetic field. The
resultant  parameters ($\hbar\omega_{c}/B$ and $\delta_{0}$) are
displayed in Figs.~\ref{Fig1}c and \ref{Fig1}f (open dots) as a
function of the energy $\hbar\omega_{c}(B)$ (in these figures some
typical error bars reflect the estimated uncertainties of the
fitting procedure). From these plots we notice that below the RBE,
$\delta_{0}$ is constant and weak being determined mainly by
$\eta$. Above the RBE there are sudden and large increases in
$\delta_{0}$ for both samples. However their behaviors are quite
different. For sample MA-2226, there are two distinct
contributions to $\delta_{0}$: one lower field component starts
very large but decreases strongly with field and a second
component which starts from almost zero but goes through a maximum
at $B$ corresponding to about 48~meV. For the time being we will
label the interaction which contributes to this peak in
$\delta_{0}$ as XP.  For sample MA-1490, only the XP contribution
to $\delta_{0}$  is observed. In this case the peak occurs at $B$
corresponding to 42~meV.

Thus our results suggest that there are two kinds of excitations
interacting with the Q2D electron gas at high fields: one which is
observed in the lower doped samples and another one labelled as XP
which is observed in both samples. Very similar results have been
obtained with another sample MA-2227 doped at the level
$N_{s}=2\times 10^{11}$~cm$^{-2}$. In this sample the first
interaction is a little bit stronger than for sample MA-2226 while
the XP peak is quite similar. We assign the first interaction to
polaronic effects since it appears to peak at fields when the CR
frequency will resonate with the GaAs LO phonon frequency (slab
mode). This assignment is consistent with the observation that
this interaction is very sensitive to the carrier concentration.
In sample MA-1490 this interaction seems to have disappeared from
$\delta_{0}$. We note that in a previous report \cite {Faugeras},
on lift-off samples more highly doped in the range $N_{s}= 7 $ to
$9\times 10^{11}$~cm$^{-2}$, only some weak interaction around the
TO of GaAs was observed. This result was later on analyzed by
Klimin \textit{et al.} \cite {Klimin} as PE with the LO phonon
screened by electrons to the point that its value becomes close to
the  TO phonon frequency. The same analysis can be applied to the
sample MA-1490 and other higher doped samples.

To mimic PE we restrict ourselves to one polaron first and, using
the FHIP model, write down $\mathrm{Im}(\Sigma(\omega))$ as \cite
{PD}:

\begin{equation}
\mathrm{Im} (\Sigma(\omega))= \omega_{LO}\frac{\omega_{0}}{\omega}\alpha
F|A_{0}(\omega)|^{1/2}e^{-R|A_{0}(\omega)|}\Theta(A_{0}(\omega))
\end{equation}
where $A_{0}(\omega)= \omega/\omega_{0}-1$, $\Theta(x)=1$ for $
x>1$ or zero otherwise. In the absence of screening $\omega_{0}
=\omega_{LO}$. If $\omega_{0}$ is known, Eq.~(2) depends on two
parameters $\alpha F$ and $R$.  The FHIP model, developed in a one
electron picture,  assumes that  electron-phonon interaction is
harmonic  instead of  Coulombic  as in the case of the
Fr\"{o}hlich interaction. The model depends on two parameters:
\textit{v} and \textit{w} in reduced units of $\omega_{LO}$. One
of them (\textit{w}) is close to 1 while
$\textit{v}^{2}-\textit{w}^{2}$ represents the force constant of
the harmonic interaction (equivalent to the Fr\"{o}hlich constant
$\alpha$). In this case
$R=(\textit{v}^{2}/\textit{w}^{2}-1)/\textit{v}$ whereas $\alpha
F$ is a global quantity which  depends on \textit{v} and
\textit{w} but should also imply corrections for the
dimensionality of the problem and for  screening effects. The
present approach can be regarded as a test  of the FHIP model when
applied to Q2D electrons under high magnetic fields.

Using this simplified model, one can fit the data on $\delta_{0}$
for the polaronic contribution. For sample MA-2226 we have to add
 contribution for the XP interaction (Fig.~\ref{Fig1}c, upper
panel) . If we assume that the XP interaction satisfies the linear
response function theory, then  $\mathrm{Im}(\Sigma(\omega))$ and
$\mathrm{Re} (\Sigma(\omega))$ are related by the
Kramers-Kr\"{o}ning (KK) relations. The resultant constraint on
the fitting process is that the  KK transformation of
$\delta_{0}(\omega)$ should reproduce the  variation of
$\hbar\omega_{c}(\omega)/B$. One has also to fit the NP effects,
$\hbar\omega_{NP}(B)/B$ versus $\hbar\omega_{NP}(B)$ with standard
models \cite {Hermann} (dashed oblique lines in Fig.~\ref{Fig1}c)
but this corresponds simply to a global shift of the curves. The
fitted functions $\hbar\omega_{c}(\omega)/B$ and
$\delta_{0}(\omega)$ are then inserted in the multilayer
dielectric model from which one can compute, for each value of  B,
the TR spectrum and compare it to the corresponding experimental
spectrum (Fig.~\ref{Fig1}b). The agreement is quite satisfactory
as one can see from that figure. However,  it is clear that the
fitting of PE is not unique for these reasons: (i) we have assumed
for sample MA-2226 that $\omega_{0} =\omega_{LO}$ neglecting,
therefore, screening effects, (ii) we have treated the problem
with one polaron instead of several polarons. Even if this effect
does not influence $\mathrm{Im}(\Sigma(\omega))$, the KK
transformation depends on \textit{all} contributions to
$\mathrm{Im}(\Sigma(\omega))$ including those at higher energies
than we can reach experimentally. The same procedure has been
applied  to simulate the data on sample MA-1490, but here we have
assumed  a weakened PE starting near the TO phonon energy of GaAs
in order to reproduce  the non-linearity of
$\hbar\omega_{c}(\omega)/B$ below the RBE (lower panel of
Fig.~\ref{Fig1}f). The choice of the corresponding fitting
parameters in Eq.~(2) is now completely arbitrary. We can also
reproduce quite well the experimental transmission spectra as
shown in Fig.~\ref{Fig1}e.

Although we do not know the value of the pre-factor $\alpha F$ in
Eq.~(2), it is instructive to compare the parameter $R$ entering
this equation. One finds  $R \approx 25$ and $18$ for samples
MA-2227 and MA-2226 respectively. These values could, of course,
be lowered if one assumes some screening of the LO phonon in
Eq.~(2) and if we include higher polarons in the analysis but they
remain, at least, an order of magnitude higher than that proposed
($R=0.04)$ for unscreened PE \cite{PD1}. These values imply that
the \textit{v} and
 \textit{w} parameters  in the FHIP model have to be smaller than
 1 even for the samples MA-2226
and MA-2227. This raises the question of the physical
 interpretation of these parameters: knowing that $\omega_{LO}$ cannot be smaller than
 $\omega_{TO}$, it seems that a more physical approach should be
 to relate in some way these parameters to the TO-LO splitting rather
 than to a single frequency like $\omega_{LO}$.

\begin{figure}
\includegraphics*[width=0.9\columnwidth]{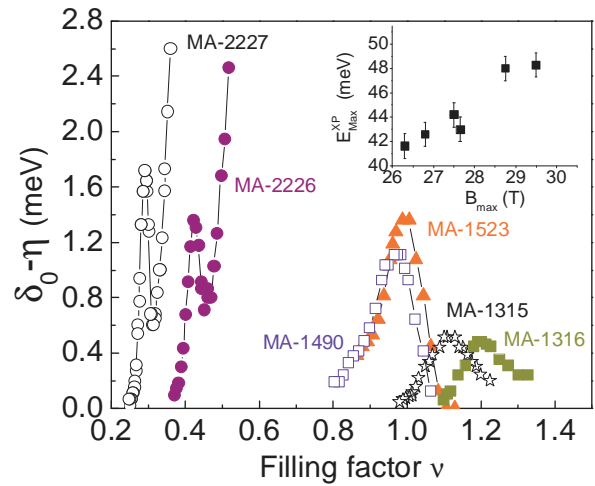}
\caption{\label{Fig2}(Color on line) Variation of $\mathrm{Im}
(\Sigma)$ with the filling factor $\nu$ for different samples:
MA-2227 with open dots ($N_{s}=2\times 10^{11}$~cm$^{-2}$),
MA-2226 with full dots ($N_{s}=3\times 10^{11}$~cm$^{-2}$),
MA-1490 with open squares ($N_{s}=6.3\times 10^{11}$~cm$^{-2}$),
MA-1523 with full triangles ($N_{s}=6.6\times 10^{11}$~cm$^{-2}$),
MA-1315 with open stars ($N_{s}=7.1\times 10^{11}$~cm$^{-2}$),
MA-1316 with full squares ($N_{s}=7.7\times 10^{11}$~cm$^{-2}$).
Insert: Variation of the characteristic energies $E_{max}^{XP}$ of
the maximum of the XP interaction with the corresponding magnetic
field $B_{max}$(see text). The corresponding error bars take into
account the non symmetric shape of the interaction.}
 \end{figure}

 We focus next on  the discussion on the possible origins of the XP
 interaction. The variation of $\mathrm{Im}(\Sigma)= \delta_{0}-\eta$ is
 plotted in Fig.~\ref{Fig2} for six different samples as a function of the
 filling factor $\nu= N_{s} \Phi_{0}/B$ where $\Phi_{0}$ is the flux
 quantum. The XP interaction which is \textit{only} observed for energies
 larger than the RBE, clearly decreases for the higher doped
 samples (MA-1315 and MA-1316) and indeed disappears completely when
 $N_{s}\geq 9\times 10^{11}$~cm$^{-2}$. This result indicates that
 the XP interaction is strongly screened by free electrons.
 The energy $E_{max}^{XP}$ corresponding to
 the maximum of this interaction decreases while  the corresponding
 value $\nu_{max}$, the value of $\nu$ where this maximum occurs, increases  or $E_{max}^{XP}$ increases with
 the  corresponding value of $B_{max}$(insert of
 Fig.~\ref{Fig2}).
 This interaction is clearly
 dissipative and in these samples of very high mobility,   is
 unlikely to be caused by impurities.
 We are then left with intrinsic mechanisms  such as electron-phonon
 interaction as a plausible explanation knowing that in these samples
 the inter-subband energy is of the order of 65 meV \cite{Faugeras,Bychkov} and cannot play a role.
 In a GaAs QW, sandwiched between AlAs layers,
 several kinds of optical phonons are
 present. The most well known ones are the confined  phonons
 (slab modes) mainly associated with mechanical motion of atoms.
 The frequency of these modes  lies inside the RBE.
 The other phonons of dielectric origin (which
 could be hence screened by free carriers) are the interface
 phonons.
  Because of the reflection symmetry of the
 GaAs QW with respect to its center, these modes are divided into
 either symmetric or anti-symmetric modes \cite{Mori}.
 The symmetric modes which are infra-red active, have been invoked in Ref.~\cite{Wang}
 to explain the ``splitting'' of the CR transition observed near
 $\hbar\omega_{TO}(AlAs)$($45$~meV). Such splitting is observed also in
 \textit{all} our samples including the very heavily doped samples (up to
 $N_{s}= 1.9\times 10^{12}$~cm$^{-2}$). As seen in Figs.~\ref{Fig1}b and \ref{Fig1}e, it can be perfectly
 reproduced by our simulation with the multi-layer dielectric model as
 a result of pure interference effects in the transmission spectra.
 We also note that the wave function of these symmetric interface modes does not
 vanish at the center of the QW (unlike the anti-symmetric modes) and therefore have
 significant overlap  with the wave function of the Q2D gas. Thus
 the idea of possible interaction between the interface modes and
 the Q2D electron gas deserves more detailed analysis. The
 solutions for these modes between a GaAs layer sandwiched between
 two AlAs layers are  given by \cite{Mori}:

\begin{equation}
\varepsilon_{AlAs}= -\varepsilon_{GaAs}\times \tanh(q_{//}L/2)
\end{equation}
 where $q_{//}$ is the wave vector of the interface electromagnetic wave
 travelling parallel to the plane of the Q2D gas. In the absence of magnetic
 field, the two solutions of Eq.~(3) lie in the Reststrahlen bands of GaAs and
 AlAs.
  But, when
 the magnetic field effect is included in $\varepsilon_{GaAs}$,
 through Eq.~(1) for instance, the results are  different: (i) depending on the parameters
 the number of solutions of Eq.~(3) can be larger than 2, (ii) some of the solutions are
 now strongly dependent on the magnetic field
  and can extend to energies higher than $\hbar\omega_{TO}(AlAs)$ as observed
 here.
 When increasing $N_{s}$, one would expect that an interaction with the Q2D electron gas will
  result in a downshift (or renormalization) of the interface mode energy in a way similar
  to what is observed for the polaronic effects in Figs.~\ref{Fig1}c and \ref{Fig1}f.
 This effect has never been investigated to our knowledge but
 certainly deserves further studies. In particular it is desirable to
 understand the mechanism of electron-phonon interaction which makes this interaction resonant
 for a specific energy $E_{max}^{XP}$.

 At present we prefer the explanation of the XP interaction in terms of interface modes
 though we cannot exclude  more speculative
 interpretations, such as, strong non-linear effects of the polaronic interaction.

 It is clear that in order to get a deeper and more definitive understanding
  into all  the interactions between the Q2D electrons and other
  elementary excitations like phonons, one should obtain
  experimental results  on lift-off samples which can provide more unambiguous
  values of the threshold energies related to the simple polaronic effect.

In conclusion, we have performed infra-red  magneto-optical
transmission measurements of a Q2D electron gas in a single
modulation-doped GaAs QW with different densities, for magnetic
fields high enough to scan the cyclotron resonance frequency
beyond the Reststrahlen band of GaAs. From the experimental
spectra, we have extracted the imaginary part of the response
function which reveals several singularities whose number and
strength depend on the carrier density. These singularities have
been attributed to electron-phonon interactions. One of these
interactions involving the LO phonon of GaAs has been treated with
a simplified version of the FHIP polaronic model. But to explain
all our results quantitatively, it is necessary to invoke a more
elaborate theory which includes effects of screening and possible
interaction with interface modes.

The GHMFL is ``Laboratoire conventionn\'{e} \`{a} l'UJF  de
Grenoble''. This work has been supported in part by the European
Commission through the Grant RITA-CT-2003-505474. G.M and P.Y.Y
acknowledge support of a grant from the France-Berkeley fund.

\bibliography{basename of .bib file}

\begin{references}

\bibitem {Devreese} J.\ T.\ Devreese, J.\ Phys.:\ Condens.\
Matter \textbf{19}, 255201 (2007) and references herein.

\bibitem {Gaal} P.\ Gaal, W.\ Kuehn, K.\ Reimann, M.\ Woerner, T.\ Elsaesser and R.\
Rey, Nature \textbf{450}, 1210 (2007).

\bibitem {Chen} Yu.\ Chen, N.\ Regnault, R.\ Ferreira, Bang-Fen.\ Zhu,  and G.\
Bastard, Phys.\ Rev.\ B \textbf{79}, 235314 (2009).

\bibitem {Wang} Y.\ J.\ Wang, H.\ A.\ Nickel, B.\ D.\ McCombe,
F.\ M.\ Peeters,  J.\ M.\ Shi, G.\ Q.\ Hai, X.\ -G.\ Wu, T.\ J.\
Eustis and W.\ Schaff, Phys.\ Rev.\ Lett.\ \textbf{79}, 3226
(1997).

\bibitem{Faugeras} C.\ Faugeras, G.\ Martinez, A.\ Riedel, R.\
Hey, K.\ J.\ Friedland and Yu.\ Bychkov, Phys.\ Rev.\ Lett.\
\textbf{92}, 107403 (2004).

\bibitem{Lee} T. D.\ Lee and D.\ Pines, Phys.\ Rev.\ \textbf{92}, 883
(1953).

\bibitem{Feynman} R.\ P.\ Feynman, Phys.\ Rev.\ \textbf{97}, 660
(1955).

\bibitem{FHIP} R.\ P.\ Feynman, R.\ W.\ Hellwarth, C.\ K.\ Iddings and P.\ M.\ Platzman,
 Phys.\ Rev.\ \textbf{127}, 1004 (1962).

\bibitem{PD} F.\ M.\ Peeters and J.\ T.\ Devreese, Phys.\ Rev.\ B \textbf{28}, 6051
(1983).



\bibitem{Bychkov} Yu.\ Bychkov, C.\ Faugeras and G.\ Martinez,
Phys.\ Rev.\ B \textbf{70}, 085306 (2004).

\bibitem{Hermann} C.\ Hermann and C.\ Weisbuch, Phys.\ Rev.\ B \textbf{15}, 823
(1977).


\bibitem{Klimin} S.\ N.\ Klimin, V.\ M.\ Fomin and J.\ T.\
Devreese, Phys.\ Rev.\ B \textbf{77}, 205311 (2008).


\bibitem{PD1} F.\ M.\ Peeters and J.\ T.\ Devreese, Phys.\ Rev.\ B \textbf{25}, 7281
(1982).


\bibitem{Mori} N.\ Mori and T.\ Ando, Phys.\ Rev.\ B \textbf{40}, 6175
(1989).





\end{references}

\end{document}